\documentclass[aps,twocolumn,superscriptaddress,altaffilletter,lengthcheck, tightenlines, showpacs,showkeys]{revtex4}

\newcommand{\al}{\alpha}

\newcommand{\D}{\Delta}
\newcommand{\ben}{\begin{eqnarray}}
\newcommand{\een}{\end{eqnarray}}
\newcommand{\be}{\begin{equation}}
\newcommand{\ee}{\end{equation}}
\newcommand{\ba}{\begin{eqnarray}}
\newcommand{\ea}{\end{eqnarray}}
\newcommand{\n}{\label}
\newcommand{\no}{\noindent}

\newcommand{\ga}{\gamma}
\newcommand{\ro}{\rho}
\newcommand{\om}{\omega}

\newcommand{\bn}{\begin{equation}\label}

\newcommand{\es}{exotic scalar field }

\usepackage[dvipdf]{epsfig}
\usepackage{color}

\begin{document}

\title{Big brake  singularity is accommodated as  an exotic quintessence field}

\author{Luis P. Chimento}\email{chimento@df.uba.ar}
\affiliation{Departamento de F\'{\i}sica, Facultad de Ciencias Exactas y Naturales,  Universidad de Buenos Aires and IFIBA, CONICET, Ciudad Universitaria, Pabell\'on I, Buenos Aires 1428 , Argentina}
\affiliation{Departamento de F\'{\i}sica, Facultad de Ciencias, Universidad del
B\'{\i}o-B\'{\i}o, Avenida Collao 1202, Casilla 5-C, Concepci\'on, Chile}
\author{Mart\'{\i}n G. Richarte}\email{martin@df.uba.ar}
\affiliation{Departamento de F\'isica, Universidade Federal do Paran\'a, Caixa Postal 19044, 81531-990 Curitiba, Brazil}
\affiliation{Departamento de F\'{\i}sica, Facultad de Ciencias Exactas y Naturales,  Universidad de Buenos Aires, Ciudad Universitaria, Pabell\'on I, Buenos Aires 1428 , Argentina}

\bibliographystyle{plain}
\begin{abstract}
We describe a big brake singularity in terms of a modified Chaplygin gas equation of state $p=(\ga_{m}-1)\rho+\al\ga_{m}\rho^{-n}$, accommodate this late-time event as an exotic quintessence model obtained from an energy-momentum tensor, and  focus on the cosmological behavior of the exotic field, its kinetic energy and the potential energy. At the background level, the exotic field does not blow up whereas its kinetic energy and potential both grow without limit  near the future singularity.  We evaluate  the classical stability of this background solution  by examining  the scalar perturbations of the metric along with the inclusion of entropy perturbation in the perturbed pressure.  Within  the Newtonian gauge,  the gravitational field approaches a constant near the singularity plus additional  regular terms.  When the  perturbed exotic field is associated with $\al>0$ the  perturbed pressure  and contrast density both diverge,  whereas the perturbed exotic field and the divergence of  the exotic field's velocity go to zero exponentially. When the  perturbed exotic field is associated with $\al<0$ the  contrast density always blows up, but the perturbed pressure can remain bounded. In addition, the perturbed exotic field and the divergence of the exotic field's velocity vanish near the big brake singularity.  We also briefly look  at the behavior of the intrinsic entropy perturbation near the singular event.
\end{abstract}
\vskip 2cm


\date{\today}
\maketitle
\section{Introduction}
One of the most challenging riddles in the last decades concerns the lack of understanding of  current state of the Universe. Supernovae data are compatible with an accelerating Universe; however, the identity of the mysterious fuel which provokes such a speeding up is currently unknown \cite{Book}.  This agent is usually dubbed dark energy and is characterized by a  negative pressure which ensures the violation of the strong energy condition. One of the missing links in the standard cosmology framework is  the lack of a fundamental  theory which can describe the main properties of dark energy at the microscopic level. So, a somewhat natural question to ask is,  what is the fundamental particle associated with dark energy?. 

Another point that should be addressed refers to the final fate of the Universe: will the universe expand forever, or will it slow down in the near future? These questions are also physically interesting and must be explored extensively at different levels. For instance, one can study what kind of final fates are compatible with the current state of the Universe. Historically speaking,  the first attempt at examining the final state of the Universe was carried out by 
 Barrow \emph{et al.}  many years ago; the authors showed the existence of  an accelerated closed FRW universe with a final singularity which exhibits an  infinite pressure \cite{ba1}.  In addition, several authors explored some accelerating universes within the dark energy scenario, endowed with a cosmological singularity in the asymptotic future  \cite{tipo1}, \cite{tipo1b}.  They also performed a full classification of the final doomday by looking at the behavior of the Hubble rate and its derivatives near the abrupt event \cite{ba1}, \cite{tipo1}, \cite{tipo1b}, \cite{tipo2}, \cite{tipo2a}, \cite{tipo2aa}, \cite{tipo2b}, \cite{tipo3}, \cite{mariam}, \cite{mariam2}, \cite{tipo4}, \cite{lip}, \cite{ba2}, \cite{mariam3}, \cite{wsingu1}, \cite{wsingu2}.  Another reliable method to understand these new cosmic singularities relies on  the existence of causal geodesics that cannot be extended to arbitrary values of their proper time (geodesic incompleteness) \cite{haw} or  the possibility of showing that  geodesic curves can be extended beyond a cosmic singularity \cite{ruth}, \cite{barrow}. A popular procedure to examine  cosmic singularities is finding the behavior of curvature invariants near the singular event provided the strength of these singularities can be determined by applying the necessary and sufficient conditions discovered by  Tipler \cite{tipler} and Kr\'olak \cite{krolak}.  


A complementary tool  for exploring the physical nature of cosmological singularities is based on the behavior of dynamical variables which enter into the field equations. Indeed,  analyzing  the blowup of  the energy density or  pressure singles out a pathological behavior that one may connect with the physical  behavior of the scale factor, the Hubble function, and its derivatives.   Hence, the idea is to find  certain physical properties  which help us to distinguish these kinds of singularities among each other \cite{tipo1b}, \cite{tipo2}, \cite{tipo2a}, \cite{tipo2aa}, \cite{tipo2b}, \cite{tipo3}, \cite{mariam}, \cite{mariam2}, \cite{tipo4}.   

There is a singularity  called a big brake which  emerges within the context of a  tachyon scalar field \cite{tipo2a}.   A weaker extension can be obtained when a dust component is included in the Friedmann equation provided the Hubble function  does not vanish  at the singularity \cite{tipo2aa}. This singularity  is reached in a finite time with a finite radius when the derivative of the  scale factor vanishes  and the acceleration term becomes unbounded \cite{tipo2a}. We are going to  study  a viable cosmological scenario where  the aforesaid singularity
can appear naturally in order to explore its physical outcome.  We present an interacting dark energy model \cite{jefe1}, \cite{jefe2}  with a  phenomenological interaction that leads to the existence of a big brake event \cite{jefe3}. We start our research by focusing on  the interacting dark energy model and its unified counterpart. In order to do that, we apply the source equation method as it seems to have the virtue (among others) that it  allows us to reconstruct the partial densities associated with dark matter and dark energy along with the total energy density in terms of the scale factor \cite{jefe1}, \cite{jefe3}. Further, we  show that the unified model is related to the modified Chaplygin gas model, which in turn reveals  the physical nature of our proposal. In order to establish an  interconnection between  the particle physics world and the cosmological setup, we find a unified description in terms of a new exotic quintessence field \cite{monica}. At the background level, we show that the unified model can be mapped into an exotic scalar field theory with the bonus that it also displays the exact behavior of the potential energy, kinetic energy, and scale factor in terms of the exotic field. The equation of motion of the exotic field is altered in relation to the standard one; however, the usual quintessence can be recovered under certain conditions. This model offers a new alternative to the well-known tachyon model which attempts to unify of  dark matter and dark energy. We introduce a covariant formulation of this new model by proposing an energy-momentum tensor which is very similar to the quintessence case. To the best of our knowledge, the exotic quintessence model cannot be (apparently) obtained from a Lagrangian.  We extend our analysis by considering the cosmological perturbation around the big brake event with the aim of determining the classical stability of these solutions.  We perform the perturbations within the Newtonian gauge because there is no residual gauge freedom and we can obtain the gravitational potential straightforwardly  by solving its master Bessel equation. Once the gravitational potential is known, we proceed to find the cosmic behavior of the perturbed exotic field by solving the $0-0$ perturbed Einstein's equation. In dealing with the pressure perturbation, we take into account two kinds of terms: one represents the adiabatic contribution, while the other accounts for the nonadiabatic pressure term which is related to the intrinsic entropy perturbation.  We provide a complete analysis of all relevant quantities,  such as the  contrast density, the divergence of the  exotic field's velocity,  and the intrinsic entropy perturbation near the singular event.
At this point, we emphasize that the study of a cosmological singularity in terms of a fundamental theory was carried out by other authors as well. Barrow and Graham have recently shown that a new ultra-weak generalized sudden singularity can be supported by a scalar field  with a simple power-law potential \cite{barrowfinale}. In fact, the appearance of a finite-time future singularity within the context of inflation  for a general noncanonical scalar-tensor theory  or  a $F (R)$ gravity model was also studied  by  Nojiri \emph{et al.} \cite{siodi1}.  It turned out that the generalized sudden singularity can be compatible with the observational data coming from Planck/BICEP2 if this event takes place at the end of inflation or afterwards \cite{siodi1}.  This singularity can be described in terms of  canonical and phantom scalar fields \cite{siodi2} or  with the help of generalized equation-of-state fluids \cite{siodi3}. 

The structure of the paper is as follows. In Sec. II we present the interacting dark energy model which exhibits the big brake event, and apply the source equation method for reconstructing the total energy density, pressure, and partial densities in terms of the scale factor. We show the connection between the interacting setup and the unified model; the latter leads to the modified Chaplygin gas model. We sketch the link between the unified model and a new quintessence theory. Section III is devoted to showing that the new exotic quintessence model can be obtained from a covariant energy-momentum tensor. In Sec. IV  we review and apply the cosmological perturbation theory to the big brake event within the Newtonian gauge. We take into account nonadiabatic pressure perturbations and obtain the behavior of the gravitational potential, contrast density, the divergence of the exotic field's velocity,  and the intrinsic entropy perturbation near the singular event. 

\section{Interacting  two-component model}
The metric of spacetime is taken to be  an isotropic and homogeneous  Friedmann-Roberston-Walker (FRW) universe with zero spatial curvature ($k=0$), 
\bn{me}
ds^2=-dt^2+a^2(t)(dx^2+dy^2+dz^2)
\ee
where $a(t)$ is the expansion scale factor. We investigate an  interacting dark sector model where the universe is filled with two perfect fluids: one serves as a matter component,  while the the other represents  a variable vacuum energy (VVE) substratum. Both fluids have energy densities $\ro_m$, $\ro_x$, pressures $p_m$, $p_x$, and are described by linear equations of state $p_{m,x}=(\ga_{m,x}-1)\ro_{m,x}$ such that the constant barotropic indeces $\ga_m$ and $\ga_x$ satisfy the condition $0<\ga_x<\ga_m$. The total energy density $\ro$  and the conservation equation  associated with this interacting two-fluid model are 
\bn{r}
\ro=\ro_m+\ro_x,
\ee
\bn{co}
\ro'=-\ga_m\ro_m-\ga_x\ro_x,
\ee
where the  prime is a derivative with respect to the scale factor, $'\equiv d/d\eta=d/3Hdt=d/d\ln{(a/a_0)^3}$, and $a_0$ is some value of reference for the scale factor. We point out that the procedure outlined does not rely on the specific cosmological equations that govern the dynamic of a homogeneous isotropic flat universe \cite{jefe3}. Differentiating Eq. (\ref{co}) and combining with Eq. (\ref{r}), we get a second-order differential equation for the energy density,  
\bn{s} 
\rho''+(\ga_m+\ga_x)\rho'+\ga_m\ga_x\rho= (\ga_m-\ga_x)[\ro_x'+\ga_x\ro_x].
\ee
Thus the VVE or the matter acts as a source of the last equation through the term $\ro_x'+\ga_x\ro_x$ or $-(\ro_m'+\ga_m\ro_m)$ after using the conservation equation (\ref{co}). In a natural way, we identify the interaction term with the terms inside the square bracket $Q=\ro_x'+\ga_x\ro_x$ and this identification splits the conservation equation (\ref{co}) into two equations. Here $Q$ produces the exchange of energy between the two fluids and we assume that it is a function of $\ro$, $\ro'$, and $\eta$. Hence  Eq. (\ref{s}) leads to the  ``source equation"  
\bn{se}
\rho''+(\ga_m+\ga_x)\rho'+\ga_m\ga_x\rho= (\ga_m-\ga_x)\,Q,
\ee
which will be a useful tool for obtaining the energy density for a given phenomenological interaction term. From the algebraic linear system of  equations (\ref{r})-(\ref{co}), we can reconstruct  $\ro_m$ and $\ro_x$ as functions of $\ro$ and its $\eta$ derivative $\ro'$ :   
\be
\n{rom}
\ro_m=-\,\frac{\ga_x\ro+\ro'}{\ga_m-\ga_x}, ~~~\ro_x=\frac{\ga_m\ro+\ro'}{\ga_m-\ga_x}.
\ee
When the  energy density $\ro$  is obtained after solving the source equation (\ref{se}) for a given interaction term $Q$, we are able to find the matter and VVE energy densities in terms of the scale factor. Comparing the total pressure $p(\ro,\ro')=-\ro-\ro'$ with the effective equation of state of the dark sector $p=(\gamma-1)\rho$, we obtain the effective conservation equation $\ro'+\ga\ro=0$, where the effective barotropic index is by $\gamma=(\ga_m\ro_m+\ga_x\ro_x)/\rho$.  

In dealing with the interacting dark sector, we propose a phenomenological interaction which is a  nonlinear combination of $\ro_m$, $\ro_x$,  and $\ro$
\bn{q}
 Q=n\,\ga_m\frac{\ro_m\ro_x}{\ro},
\ee
and whose physical motivation was explored in detail in our previous article \cite{jefe3}. Taking $\ga_x = 0$ and leaving $\ga_m$ as a free parameter in  Eq. (\ref{rom}), we obtain the matter energy density and the VVE density $\ro_m=-\ro'/\ga_m$, $\ro_x=\ro+\ro'/\ga_m$ as functions of $\ro$ and $\ro'$. Replacing the latter equations in  the interaction term (\ref{q}), we see  that the source equation (\ref{se}) is considerably simplified,
\bn{sef} 
\ro\rho''+\ga_m(n+1)\ro\rho'+n\ro'\,^2 =0.
\ee

We will find the  general solution of the source (\ref{sef}) from a nonlinear superposition of the two basis solutions of a second-order linear differential equation by changing to the variable $ x=\ro^{n+1}$,
\bn{x''}
x''+\ga_m(n+1)x'=0.
\ee
Its first integral and the energy density are given by
\bn{r'}
\ro'=-\ga_m(\ro+\al\ro^{-n}),
\ee 
\bn{rf}
\ro=\left\{\al\left[-1+ \left(\frac{a_s}{a}\right)^{3\ga_m(n+1)}\right]\right\}^{1/(n+1)},
\ee
where $\al$ is one of the integration constants while the other was set so that the square bracket in Eq. (\ref{rf}) vanishes as the scale factor reaches the finite value $a_s$. This is a necessary condition to achieve a singularity.  In fact, at $a_s=a(t_s)$ the energy density (\ref{rf}), the pressure $p$, the barotropic index $\ga$, or the acceleration $\ddot a$, vanishes or diverges at a finite  time $t_s$, as can be seen from the following expressions:
\bn{pcha}
p=(\ga_{m}-1)\rho+ \al\ga_{m}\rho^{-n},
\ee
\bn{gaf}
\ga=\ga_m\left[1+\al\ro^{-n-1}\right],
\ee
\bn{a..f}
\frac{\ddot a}{a}=-\frac{1}{6}(3\ga_m-2)\ro-\frac{\al\ga_m}{2}\ro^{-n},
\ee
\bn{rmxff}
\ro_m=\ro+\frac{\al}{\ro^n}, \qquad \ro_x=-\frac{\al}{\ro^n}.
\ee
As a by-product of the nonlinear interaction (\ref{q}), we  are led to an effective one-fluid model with a modified Chaplygin gas equation of state [ Eq. (\ref{pcha})] \cite{jefe3}. Finally, the  $\eta$ derivative energy density (\ref{rf}) and the effective barotropic index expressed as functions of the scale factor are 
\bn{r'f}
\ro'=-\al\ga_m\left(\frac{a_s}{a}\right)^{3\ga_m(n+1)}\ro^{-n},
\ee
\bn{gaa}
\ga=\frac{\ga_m}{1-\left(\frac{a}{a_s}\right)^{3\ga_m(n+1)}}\,.
\ee
Note that the barotropic index always diverges ($\ga\to\infty$) in the  $a\to a_s$ limit   for any value of $n$. The remaining quantities- namely the pressure, the acceleration, and the matter and VVE densities- can be obtained as functions of the scale factor by combining Eqs. (\ref{pcha}), (\ref{a..f}), and  (\ref{rmxff})  along with Eq.  (\ref{rf}). 

Now, we are in a position to show how the interacting dark sector can be mapped into an exotic scalar field scenario. Our idea is to provide a physical interpretation of the emergence of a sudden future (big brake) singularity by linking  a phenomenological  two-interacting fluids model with a new  scalar field model.  We  will describe the model in terms of the kinetic variable $\dot\phi^2$ of the \es and the potential variable $V=V(\phi)$ instead of the dark matter and dark energy densities $\ro_m$ and $\ro_x$.  

Our starting point is to note that the conservation equation (\ref{co}) can be written as  $\ro'=-(\ga_m\ro_m+\ga_x\ro_x)=-(\ro+p)=2\dot H$, so it  suggests a natural identification  with the kinetic energy in terms of the first derivative of the total energy density,
\bn{i}
\dot\phi^2=-\ro',
\ee
which implies  that the energy density and pressure of the interacting two-fluid mixture as well as the matter energy density, 
\bn{re}
\ro=\frac{\dot\phi^2}{\ga_m}+\frac{\ga_m-\ga_x}{\ga_m}\,\ro_x,
\ee
\bn{pe} 
p=\frac{(\ga_m-1)}{\ga_m}\,\dot\phi^2-\frac{\ga_m-\ga_x}{\ga_m}\,\ro_x,
\ee
reproduce the relation 
\bn{r+p}
\ro+p=\dot\phi^2
\ee
of the quintessence scalar field. Substituting the energy density (\ref{re}) into the conservation equation (\ref{co}), we find the exotic scalar field equation for $\phi$ generalizing the Klein-Gordon equation \cite{monica},
\bn{kg}
\ddot\phi+\frac{3\ga_m}{2}H\dot\phi+\frac{(\ga_m-\ga_x)\dot\ro_x}{2\dot\phi}=0,
\ee
which shows that the  conservation equation (\ref{co}) and source equation (\ref{se}) both have the same kind of  physical information. In order to go from  the dark energy densities $\ro_m$, $\ro_x$ to the variables characterizing the \es $\dot\phi^2$ and $V(\phi)$, we  identify  the potential $V(\phi)$  with the VVE density 
\bn{v}
V(\phi)=\ro_x.
\ee

Coming back to our model, in which the barotropic index of the VVE is $\ga_x=0$, Eqs. (\ref{re}) and (\ref{pe}) can be recast as
\bn{r0}
\ro=\frac{\dot\phi^2}{\ga_m}+V(\phi),
\ee
\bn{p0} 
p=\frac{(\ga_m-1)}{\ga_m}\,\dot\phi^2-V(\phi),
\ee
while the dark matter energy density and the exotic scalar field equation become
\bn{rm0} 
\ro_m=\frac{\dot\phi^2}{\ga_m},
\ee
\bn{kg0}
\ddot\phi+\frac{3\ga_m}{2}H\dot\phi+\frac{\ga_m}{2}\,V'=0,  
\ee
where $V'=dV/d\phi$. In the particular case that the interacting dark components are associated with a scalar field in the form $\dot\phi^2=2\rho_{m}$ and $V(\phi)=\rho_{x}$, with equations of state $p_{m}=\rho_{m}$ and $p_{x}=-\rho_{x}$ (namely, stiff matter and VVE, respectively), the exotic scalar field becomes the quintessence field and the energy-momentum tensor conservation of the dark sector (as a whole) reduces to the Klein-Gordon equation. For any other interacting two-fluid mixture, we are led to an exotic quintessence field $\phi$. 

By combining  the dark matter energy density (\ref{rmxff}) with the energy density (\ref{rf}) and Eq. (\ref{rm0}), we obtain the exotic kinetic energy $\dot\phi^2/\ga_m$ as a functions of the scale factor, so we have 
\bn{k}
(\phi\,')^2=\frac{\ga_m}{3}\frac{\left(\frac{a_s}{a}\right)^{3\ga_m(n+1)}}{\left(\frac{a_s}{a}\right)^{3\ga_m(n+1)}-1}.
\ee
Assuming that the evolution of the homogeneous and isotropic flat universe is described by the Einstein field equations and the dynamics of the effective one-fluid model is governed by the corresponding Friedmann constraint, $3H^2=\ro$, we integrate  Eq. (\ref{k}) and find two sets of solutions. We have two cases to explore, depending on the sign of $\alpha$.

 \begin{itemize}
   \item $\al>0$
 \end{itemize}
Under this condition, we find that the potential and kinetic energies involves hyperbolic functions:
\vskip .2cm
\no\bn{v2} 
V=-\al\left[\al\sinh^2{\om \D\phi}\right]^{-n/(n+1)},
\ee
\bn{k2}
\dot\phi^2=\ga_m\ro\coth^2{\om\D\phi},
\ee
\bn{r2}
\ro=\left[\al\sinh^2{\om \D\phi}\right]^{1/(n+1)},
\ee
\bn{p2}
p=(\ga_m\coth^2{\om\D\phi}-1)\,\ro,
\ee
\bn{g2}
\ga=\ga_m\coth^2{\om\D\phi},
\ee
\bn{a2}
a=a_s\left[\cosh^2{\om\D\phi}\right]^{-1/3\ga_m(n+1)},
\ee
\bn{a..2}
\frac{\ddot a}{a}=\frac{1}{2}\left[\frac{2}{3}-\ga_m\coth^2{\om\D\phi}\right ],
\ee
\bn{om}
\om=-\frac{\sqrt{3\ga_m}}{2}(n+1),   
\ee
where $\D\phi=\phi-\phi_s$ and $\phi_s$ is an integration constant.
\begin{itemize}
   \item $\al<0$
 \end{itemize}
Under this condition, we find that the potential and kinetic energies involves trigonometric functions:
\no\bn{v3} 
V=-\al\left[-\al\sin^2{\om \D\phi}\right]^{-n/(n+1)},
\ee
\bn{k3}
\dot\phi^2=-\ga_m\ro\cot^2{\om\D\phi},
\ee
\bn{r3}
\ro=\left[-\al\sin^2{\om \D\phi}\right]^{1/(n+1)}.
\ee
\bn{p3}
p=(-\ga_m\cot^2{\om\D\phi}-1)\,\ro,
\ee
\bn{g3}
\ga=-\ga_m\cot^2{\om\D\phi},
\ee
\bn{a23}
a=a_s\left[\cos^2{\om\D\phi}\right]^{-1/3\ga_m(n+1)},
\ee
\bn{a..3}
\frac{\ddot a}{a}=\frac{1}{2}\left[\frac{2}{3}+\ga_m\cot^2{\om\D\phi}\right ].
\ee

In both cases, the scale factor reaches its  finite value $a_s$  and the \es also takes a finite value $\phi_s=\phi(t_s)$ where is  $t_s$ the cosmological time  where the finite-time future singularity occurs. In other words, the case $\al<0$ is related to the case  $\al>0$ by means of a Wick rotation in the exotic quintessence field, as can be noticed from the next transformations: $\cos(i\om\D\phi)= \cosh (\om\D\phi)$ and $\sin(i\om\D\phi)=i \sinh (\om\D\phi)$.



\section {Covariant approach}
So far, we have shown  the appearance of a modified Chaplygin equation of state associated to the effective fluid of two interacting dark components. Further, we demonstrated that such an effective fluid can be mapped into an exotic scalar field theory by proposing a natural identification. Our next task is to provide a covariant energy-momentum tensor associated to  the exotic quintessence model. In doing so, we propose that the energy-momentum tensor can be recast as
$$
T_{\mu\nu}=\nabla_{\mu}\phi\nabla_{\nu}\phi-\frac{(\ga_m -1)}{\ga_m}\, g_{\mu\nu}\nabla^{\alpha}\phi\nabla_{\alpha}\phi 
$$
\bn{tmn}
-\frac{(\ga_m -\ga_x)}{\ga_m}\, g_{\mu\nu}V(\phi,\nabla_\al\phi),
\ee
where the exotic scalar field $\phi$ is driven by the  generalized potential energy $V(\phi,\nabla_\al\phi)$. This is reminiscent of the situation  in generalized scalar-quintessence-type theories.
 Now, taking the covariant derivative of the energy-momentum tensor (\ref{tmn}) and projecting it along the direction of $\nabla^{\nu}\phi$, we obtain   the $\phi$ equation of motion
$$
\nabla^{\alpha}\nabla_{\alpha}\phi -\frac{(\ga_m -\ga_x)}{\ga_m}\, \frac{\nabla^\al\phi\nabla_\al V}{\nabla^{\alpha}\phi\nabla_{\alpha}\phi}
$$
\bn{exf}
+ \frac{(2-\ga_m)}{\ga_m}\,\frac{(\nabla^{\nu}\phi\nabla^{\mu}\phi).(\nabla_{\nu}\nabla_{\mu}\phi)}{\nabla^{\alpha}\phi\nabla_{\alpha}\phi}=0.
\ee
Hence, the equation of motion of the \es differs substantially from the standard  quintessence field provided the kinetic energy contributions which appear in the last two terms. 

From the  FRW background (\ref{me}) and energy-momentum tensor (\ref{tmn}) we can read off the energy density and pressure of the homogeneous exotic field as $-T^{0}_{0}=\ro$ and  $T^{i}_{j}=p\delta^{i}_{j}$.  We notice  that the generalized potential plays the same role as the dark energy density $\ro_x$. In particular, for a mix of matter and VVE $(\ga_x=0)$, and an \es driven by a potential $V=V(\phi)$ depending only on $\phi$, we reproduce  Eqs. (\ref{re}), (\ref{pe}), and  (\ref{kg}). Also, from these equations we note that the frictional term in the \es equation differs from the standard case  as well  as the potential term. 

From the physical point of view, it is appropriate to point out certain caveats of our model. One way to present this issue is by comparing our  interacting two-component model established on the exotic quintessence field \cite{monica} with the well-known big brake scenario based on a tachyon field \cite{tipo2a}. Even though  we have  presented the new exotic field from the standpoint of a covariant model, we have not  been able to obtain its Lagrangian representation.  This fact does not mean that such a Lagrangian description does not exist, given that the exotic quintessence field has not been sufficiently investigated in the literature, and it appears to be not as simple as the tachyon model \cite {tipo2a}. Our work is a contribution aimed  at understanding  the above problem.

Also,  we should take into account that our interacting two-component model includes a matter component that is not necessarily dust because in general the barotropic index $\ga_m$ is not equal to the unity, indicating  that we are dealing with warm dark matter. Of course, we can take $\ga_m$ arbitrarily close to unity. Anyway, we should stress that the effective barotropic index (\ref{gaa}) behaves as dust ($\ga=1$) when the scale factor takes the value 
\bn{g=1}
a_{dust}= a_s(1-\ga_m)^{1/3\ga_m(n+1)},
\ee
meaning that perhaps a dust component should not be too significant in the construction of the exotic quintessence scenario.


\section{Perturbation equations}
In order to  examine the physical consequences introduced by this model,  we will analyze the behavior of cosmological perturbations around the big brake singularity in both of  the cases mentioned above. The main reason  for studying  this issue is to extract the behavior of the perturbed exotic field and its potential, as well as the behavior of the gravitational potential, and by doing so we will be able to explore the classical stability of these solutions near the singularity.

In dealing with  the cosmic perturbations around the singular event, we will follow the notation of  Ma and Bertschinger \cite{MaBer}.  We will restrict our analysis to the Newtonian gauge provided  we are interested in the scalar mode of the metric perturbations; namely, vector and tensor  modes are neglected from the beginning. However, we want to emphasize that the classical stability of another type of singularities (sudden-type)  were examined by  Barrow and Lip a few years ago \cite{BaLip}, including an analysis of the vector and tensor modes using a gauge-invariant formalism developed by Bardeen. In that work, the authors performed a generic analysis  for perfect fluids that included adiabatic pressure perturbations  and did not make reference to any scalar field theory.

One of the main reasons for choosing the Newtonian gauge is that the physical observers are attached to  the unperturbed metric so they measure both the gravitational and the velocity fields. Besides, this metric tensor becomes diagonal and this simplifies calculations. Moreover, the equations have a simple physical interpretation for perturbations inside the horizon and there is no residual gauge freedom.  The perturbations of the FRW metric  in the Newtonian gauge read
\bn{Perm}
ds^2=a^2(\tau)\left\lbrace  -d\tau^2 (1+2\Psi)+(1-2\Psi)\delta_{ij}dx^i dx^j \right\rbrace,
\ee
where $\tau$ stands for the conformal time and $\Psi$ represents the analog of the Newtonian gravitational potential. Eq. (\ref{Perm}) tells us that the we can neglect shear perturbation. To linear order, the energy-momentum tensor is given by 
\bn{Pertem1}
T^{0}_{0}= - (\rho + \delta \rho),
\ee 
\bn{Pertem2}
T^{0}_{i}= (\rho + p)v_{i}=-T^{i}_{0},
\ee            
\bn{Pertem3}
T^{i}_{j}=  (p + \delta p) \delta^{i}_{j},
\ee 
where the velocity perturbation is defined as $v^i \equiv dx^i/d\tau$.  The first-order perturbed Einstein equations  can be recast as
\bn{PertEq1}
k^2 \Psi + 3 {\cal H} (\dot{\Psi}+{\cal H}\Psi)=\frac{a^2}{2}\delta T^{0}_{0},
\ee 
\bn{PertEq2}
k^2(\dot{\Psi}+{\cal H}\Psi)=\frac{a^2}{2}(\rho + p)\theta,
\ee            
\bn{PertEq3}
\ddot{\Psi}+ 3{\cal H}\dot{\Psi}+(2\frac{\ddot{a}}{a}-{\cal H}^2)\Psi=\frac{a^2}{6} \delta T^{i}_{i}.
\ee 
Above  we defined  $\theta=ik^l v_l=\nabla . \bar{v}$ and identified $(\rho + p)\theta=ik^l \delta T^{0}_{l}$,  while the conformal Hubble parameter is given by ${\cal H}=\dot{a}/a$. Besides, the perturbed part of  the energy-momentum conservation gives
\bn{PertCE1}
\dot{\delta}+  3{\cal H}\left( c^{2}_{s} -w(\rho)\right)  \delta=-\frac{\rho+p}{\rho}\left(\theta -3\dot{\Psi} \right),
\ee            
\bn{PertCE2}
\dot{\theta} +{\cal H}(1-3c^{2}_{s})\theta = \frac{k^2 c^{2}_{s} \delta}{1+ w(\rho)}+ k^2 \Psi,
\ee 
where $\delta=\delta\rho/\rho$, $w(\rho)=p(\rho)/\rho$, and $c^{2}_{s}=\delta p/\delta \rho$ stands for the speed sound of the scalar field.  Let us  take a closer look at the pressure perturbations for the exotic scalar field. As is well known,  scalar fields  can generate entropic perturbations,  implying that $\delta p|_{\rm rf}=c^{2}_{s~ {\rm rf }}\delta \rho|_{\rm rf}$ only holds in the rest frame of the exotic scalar field, where the perturbed energy momentum looks diagonal for a comoving  observer.
Thus,  we can interpret the quantity $c^{2}_{s~ {\rm rf }}$ as the speed  at which  pressure fluctuations propagate. However, we notice that for any other frame the link between pressure and density breaks down. In particular, this means that the adiabatic sound speed $c^{2}_{a}= \dot{p}/\dot{\rho}$  is  not equal to the sound speed of the scalar field. In order to establish  a general relation between pressure perturbations and density perturbations, we must take into account that the pressure perturbations have two kinds of contributions, namely, adiabatic and entropy components. For any given frame,  the intrinsic entropy perturbation of  matter is a gauge-invariant variable $\Gamma$:
\bn{Ga}
w\Gamma\equiv\frac{\delta p}{p}- \frac{c^{2}_{a}}{w}\frac{\delta \rho}{\rho}=\frac{\dot{p}}{\rho}\left( \frac{\delta p}{\dot{p}}-\frac{\delta\rho}{\dot{\rho}}\right). 
\ee 
Notice that $\Gamma$ is a dimensionless quantity. Here, we interpret the intrinsic entropy perturbation as a displacement between hypersurfaces of uniform  pressure and uniform energy density. Using a gauge transformation between the density perturbation $\delta \rho$ and  the density perturbation in the rest frame $\delta \rho|_{\rm rf}$,
\bn{gau}
\delta \rho|_{\rm rf}=\delta \rho + 3{\cal H}\frac{(p+\rho)\theta}{k^2}
\ee
we can determine  the pressure perturbation in a general frame in terms of the rest-frame sound speed,
\bn{link}
\delta p =c^{2}_{s~ {\rm rf }}\delta \rho + 3{\cal H}\frac{(p+\rho)\theta}{k^2}[c^{2}_{s~ {\rm rf }}-c^{2}_{a}]. 
\ee 
It is worth noting that we have to use Eq. (\ref{link}) to obtain  $c^{2}_{s~ {\rm rf }}$. As stated earlier, in a general frame, the pressure perturbation involves a density perturbation along with  an additional velocity term which accounts for the entropy generation.

Now,  using  Eqs. (\ref{tmn})  and (\ref{Perm}) we calculate the perturbed density and perturbed pressure associated with the exotic quintessence model as 
\bn{dex}
\delta \rho=\frac{2(1-\ga)}{a^2}\left(\dot{\delta\phi}\dot{\phi}-\Psi\dot{\phi}^2\right)+ \epsilon V_{, \phi} \delta\phi,
\ee
\bn{prex}
\delta p=\frac{2\ga}{a^2}\left(\dot{\delta\phi}\dot{\phi}-\Psi\dot{\phi}^2\right)- \epsilon V_{, \phi} \delta\phi,
\ee
where $\gamma=(\ga_m -1)/\ga_m$ and $\epsilon=(\ga_m -\ga_x)/\ga_m$. Equations (\ref{dex}) and (\ref{prex}) are quite similar to  those expressions associated with the standard quintessence model. Indeed, the standard model can be recovered by demanding that the coefficients which  appear in the kinetic energy are both equal; this choice leads to $\ga_{m}=2$ and $\ga_{x}=0$, so our scheme contains the quintessence model.  The main difference in relation to the standard model is that now the kinetic energy term does not appear with the same coefficient in the perturbed density and perturbed pressure. 

The adiabatic sound speed for the exotic scalar field can be written as 
\bn{assound}
c^{2}_{a}=\frac{\frac{3\ga{\cal H}\dot{\phi}^2}{(1-\ga)a^2}+ \epsilon \frac{V_{, \phi}\dot{\phi}}{\ga-1}}{3\frac{{\cal H}\dot{\phi}^2}{a^2}},
\ee
where we have used the equation of motion (\ref{kg}) to eliminate the second derivative of $\phi$ in Eq. (\ref{assound}). 

In order to obtain the rest-frame  sound speed $c^{2}_{s~ {\rm rf }}$ of the exotic scalar field,  we use Eqs. (\ref{re}), (\ref{pe}), (\ref{link}), (\ref{dex}), (\ref{prex}), and (\ref{assound}). After some manipulation, we find that the rest-frame sound speed can be recast as 
\bn{ssound}
c^{2}_{s~ {\rm rf }}=\frac{\ga}{(1-\ga)}.
\ee
To avoid the propagation of  non-casual perturbations, we must  demand that  $0<c^{2}_{s~ {\rm rf }}\leq 1$, which implies that $0<\ga \leq 1/2$. The aforesaid constraint can be alternatively translated as $1<\ga_{m} \leq 2$. One comment might be in order here: the adiabatic  sound speed (\ref{assound}) is a background quantity which should be clearly distinguished from the speed of sound, which is essentially a perturbative quantity.

Our next task is to determine a master equation for the gravitational potential. Combining Eqs. (\ref{Pertem1})- (\ref{Pertem3}) with  Eqs. (\ref{PertEq1}) - (\ref{PertEq3}),  we find 
\bn{MasterE}
\ddot{\Psi}+3{\cal H}(1+3c^{2}_{a})\dot{\Psi}+ \left\lbrace 2 \dot{{\cal H}} + k^2 c^{2}_{s {\rm rf }} + (3c^{2}_{a}+1){\cal H}^2\right\rbrace=0. 
\ee 
In this way, the master equation for the gravitational potential (\ref{MasterE})  has time-dependent coefficients that involve only background quantities. As  the pressure perturbation contains adiabatic and entropy contributions for the exotic scalar field, the master equation for the gravitational potential $\Psi$ exhibits both contributions encoded in the distinctive roles played by $c^{2}_{a}$ and $c^{2}_{s~ {\rm rf }}$.

After solving  Eq. (\ref{MasterE}), we must use the constraint (\ref{PertEq1}) again to derive the leading term of the density perturbation near the singularity; and by doing so, we arrive at a first-order differential equation for the perturbed field $\delta \phi$,
\bn{MasCon}
(1-\ga)\dot{\delta\phi}+ \frac{a^2}{2}\epsilon \frac{V_{, \phi}}{\dot{\phi}} \delta\phi=(1-\ga)\Psi\dot{\phi}- \frac{[k^2\Psi + 3 {\cal H} (\dot{\Psi}+{\cal H}\Psi)]}{\dot{\phi}},
\ee 
where the rhs of Eq. (\ref{MasCon}) is known. Another method is to solve  the perturbed equation of motion for the exotic field, which can be obtained from Eq. (\ref{PertCE1}) or by explicitly perturbing the equation of motion (\ref{exf}) around its background solution as $\phi \rightarrow \phi+ \delta \phi $, 
\[\ddot{\delta\phi}+\frac{2\ga +1}{2(1-\ga)}{\cal H}\dot{\delta\phi}+\frac{a^2}{2(1-\ga)}\left(\frac{k^2}{a^2} + \epsilon V_{, \phi\phi} \right)\delta\phi=  \]
\bn{PerKG}
\frac{5-2\ga}{2(1-\ga)}\dot{\Psi}\dot{\phi} - \frac{\epsilon}{(1-\ga)}\Psi V_{, \phi}.
\ee 
Eq. (\ref{PerKG}) suggests that  the potential energy is a function of the exotic scalar field only.  Considering the off-diagonal component of the perturbed energy-momentum tensor associated with the exotic quintessence field $T^{i}_{0}=\dot{\phi}\partial^{i}\delta\phi/a^{2}=(\rho+p)v^{i}$, we can derive $v^{i}=\partial^{i}\delta\phi/(\dot{\phi}a^{2})$ and therefore  the divergence of the exotic field's velocity is given by  $\theta=-k^2\delta\phi/(\dot{\phi}a^{2})$. In short, solving Eq. (\ref{MasCon}) also leads us to the functional form of $\theta$ while the contrast density is simply obtained  from its definition, $\delta=\delta\rho/\rho$.

At this point, it is essential to use another time variable called $T$ to analyze the leading behavior of different contributions which appear in Eq. (\ref{MasterE}). To be more precise, plugging Eq.  (\ref{a2}) or Eq. (\ref{a23}) into the Friedmann constraint leads to $\phi= \phi_{s}+ \omega (\pm \al)^{1/2}T^{\nu/2}$, depending on the sign of $\al$, where $\nu=(2n+2)/(2n+1)$. Here, the new cosmic time is a linear redefinition of the standard cosmic time $t$, so it can be written as $T=c_{0} (t_{s}-t)$ with $c_{0}=\sqrt{3}\al \ga_{m}/2(\nu -1)$. Using this time variable, one can show that the leading terms of the total density, pressure, and adiabatic sound speed  as $T \rightarrow 0$ are given by  
\bn{det1}
\rho \simeq T^{2(\nu-1)},
\ee
\bn{pet1}
p \simeq \alpha \ga_{m}T^{(\nu-2)},
\ee
\bn{c2a}
c^{2}_{a} \simeq \frac{\al \ga_{m}(\nu-2)}{2(\nu-1)}T^{2(\nu-1)}.
\ee
After changing of time variable in Eq. (\ref{MasterE}), going from the conformal time to the new cosmic time $T$, plugging Eqs. (\ref{det1})-(\ref{c2a}) into Eq. (\ref{MasterE}), and retaining the leading terms, we arrive at a much simpler master equation, 
\bn{MasterE2}
\ddot{\Psi}+ \xi_{1}T^{-1}\dot{\Psi}+ \xi_{2}T^{(\nu-2)}\Psi=0, 
\ee 
where $\xi_{1}=\kappa(2-\nu)/3$ and  $\xi_{2}=-2\nu(\nu-1)/3\al\ga_{m}$ are two constants while $\kappa=\pm 1$. It is essential to bear in mind that a big brake singularity is achieved for $1<\nu<2$, corresponding to a nonzero scale factor $0<a(t_{s})<\infty$, vanishing energy density, and infinite pressure,  leading to a cosmic scenario where   at finite time the acceleration diverges, $\ddot{a}(t_{s})=-\infty$. 

It is convenient to  define $T=(y/c)^s$ (where $s$ and $c$ are  constants to determined later) \cite{BaLip}, and insert this relation into the master equation (\ref{MasterE2}),
\bn{MasterE3}
y^2\Psi''+ (1-s + s\xi_{1}) y\Psi'+ s^2\xi_{2}(\frac{y}{c})^{s\nu}\Psi=0.  
\ee 
In order to show that Eq. (\ref{MasterE3}) is analogous to a Bessel equation we need to propose a useful parametrization. Hence,  we insert $\Psi= y^{m}{\cal P}(y)$ into Eq. (\ref{MasterE3}), which  now reads 
\bn{MasterE4}
y^2{\cal P}''+ (1+2m-s + s\xi_{1}) y{\cal P}'+ [m(m-s-s\xi_{1})+\frac{s^2\xi_{2}}{{c}^{s\nu}}y^{s\nu}]{\cal P}=0.  
\ee 
We fix our parametrization by choosing $s\nu=2$, $2m=s(1-\xi_{1})$, and $c=s\sqrt{\xi_{2}}$ \cite{BaLip}. The latter choice leads to 
\bn{MasterE5}
y^2{\cal P}''+  y{\cal P}'+ [m^2-y^{2}]{\cal P}=0,  
\ee 
whose solution involves the Bessel functions of the first and second kind, called $J_{m}$ and $Y_{m}$, respectively, 
\bn{solu}
{\cal P}=A J_{m}(y)+BY_{m}(y),  
\ee 
where $A$ and $B$ are two integration constants. By keeping the leading term of $J_{m}$ and $Y_{m}$  as $y \rightarrow 0$, we arrive at $\Psi \simeq A y^{2m} + B$ \cite{BaLip}. Coming back to our original time variable, the gravitational potential reads 
\bn{solu1}
\Psi=A T^{\kappa \xi_{\pm}}+B.  
\ee 
If $\kappa=+1$  then we have $\xi_{+}=(\nu+1)/3$, while the choice $\kappa=-1$ leads us to $\xi_{-}=(5-\nu)/3$. Notice that the gravitational potential does not depend on the wave number $k$ explicitly. 

Let us first analyze the case with $\kappa=-1$,  $\xi_{-}=(5-\nu)/3$, and $\al>0$. Replacing (\ref{solu1}) into (\ref{PertEq1}), we obtain the contrast density
\bn{solud2}
\delta \rho=\frac{-2k^{2}B}{a^{2}_{s}}+ \frac{2B}{\sqrt{3}} T^{3(\nu -1)}.  
\ee 
Also, from the equation above  we derive that $\delta \propto T^{2(1-\nu)}$, which becomes divergent as $T \rightarrow 0$. For purposes of comparison, let us also give the leading term in the pressure perturbation, 
\[\delta p=\frac{-2k^{2}Bc^{2}_{s~ {\rm rf }}}{a^{2}_{s}}+  \]
\bn{solup2}
 \frac{\sqrt{3}c_{0}A \al \ga_{m}(2-\nu)}{(\nu-1)} T^{-\frac{(\nu +1)}{3}}+ \frac{B \al \ga_{m}(2-\nu)}{(\nu-1)} T^{(\nu -2)}.  
\ee 
Thus we have two kinds of power-law terms in Eq. (\ref{solup2}) provided both diverge for $\nu \in (1, 2)$ but in a different way  for each one; namely, for $\nu \in (1, \nu_{c})$ one term dominates over the other one (with  $\nu_{c}=5/4$), while for $\nu \in ( \nu_{c}, 2)$ the situation is reversed. Equation (\ref{solup2}) tells us that the divergent parts of pressure perturbations are associated with the intrinsic entropy perturbation. 

Inserting Eqs. (\ref{v2}), (\ref{det1}), and  (\ref{solud2}) into Eq. (\ref{MasCon}) and keeping only the leading term, we obtain a first-order differential equation for the exotic field perturbation,
\bn{caso1}
\dot{\delta\phi} + m_{1} T^{(\nu-4)}\delta\phi = m_{2}T^{\frac{\nu-2}{2}}, 
\ee 
with $m_{1}<0$ and $m_{2}>0$ in the case of $B>0$. The perturbed exotic field is 
\bn{caso1a}
\frac{\delta\phi}{\delta \phi_{0}}=e^{\frac{|m_{1}|}{\nu -3}T^{(\nu-3)}}\left(1 -T^{-\nu}\frac{D_{\nu}}{\delta \phi_{0}}\Gamma \left[ \frac{\nu}{2(\nu -3)},\frac{|m_{1}|T^{(\nu-3)}}{\nu -3}\right]     \right), 
\ee 
where  $\Gamma [d, z]$ is the incomplete gamma function whereas $D_{\nu}$ is a constant which depends on $\nu$, $|m_{1}|$, and $|m_{2}|$. From Eq. (\ref{caso1a}), we obtain that the  perturbed exotic field is exponentially suppressed as $T \rightarrow 0$, so near the big brake the perturbation decays to zero. F  urther, the divergence $\theta \propto T^{(2-\nu)/2}\delta \phi$ also goes to zero near the singularity. Although the potential energy and kinetic energy at the background level behave as a negative power law near the singularity, the perturbed exotic field decays exponentially to zero,  assuring  the classical stability of this solution. 

We now deal with the other  case corresponding to the options $\kappa=+1$, $\xi_{+}=(\nu+1)/3$,   and $\al<0$.  The contrast density can be written as
\bn{solud3}
\delta \rho=\frac{-2k^{2}B}{a^{2}_{s}}+ \frac{A\al\ga_{m}}{\sqrt{3}(\nu-1)} T^{\frac{(7\nu -8)}{3}},
\ee 
and therefore the contrast is  
\bn{solud4}
\delta=\frac{-2k^{2}B}{a^{2}_{s}}T^{(2-\nu)}+ \frac{A\al\ga_{m}(\nu+1)}{2\sqrt{3}(\nu-1)} T^{\frac{(\nu -2)}{3}}.
\ee 
Then, the density perturbation vanishes for $\nu \in (8/7, 2)$ but becomes divergent when $\nu \in (1, 8/7)$. Equation (\ref{solud4}) tells us that the contrast variable blows up at the big-brake event. We turn our attention to pressure perturbations,

\bn{pertup}
\delta p=\frac{-2k^{2}Bc^{2}_{s~ {\rm rf }}}{a^{2}_{s}}- \frac{2c_{0}A (1+\nu)}{\sqrt{3}} T^{\frac{(\nu -5)}{3}}.
\ee 
Since $\nu \in (1,2)$,  we obtain that the pressure perturbations (\ref{pertup}) always diverge as $T \rightarrow 0$.  

Inserting  Eqs.
(\ref{v3}), (\ref{det1}), and (\ref{solud3}) into Eq. (\ref{MasCon}) and keeping only the leading term, we  arrive at a first-order differential equation for the exotic field perturbation,
\bn{caso2}
\dot{\delta\phi} -|m_{3}| T^{(\nu-4)}\delta\phi = i |m_{4}|T^{\frac{\nu-2}{2}}, 
\ee 
where $m_{3}$ and $m_{4}$ are two constants which depend  of $\nu$, $\al$, and $\ga_{m}$ only. The perturbed exotic field is given by 
\bn{caso2a}
\frac{\delta\phi}{\delta \phi_{0}}=e^{\frac{|m_{3}|}{\nu -3}T^{(\nu-3)}}\left(1 -T^{-\nu}\frac{E_{\nu}}{\delta \phi_{0}}\Gamma \left[ \frac{\nu}{2(\nu -3)},\frac{|m_{3}|T^{(\nu-3)}}{\nu -3}\right]     \right), 
\ee 
where   $E_{\nu}$ is a constant. From Eq. (\ref{caso2a}), we obtain that the perturbed exotic field is again exponentially suppressed as $T \rightarrow 0$,  implying that   $\theta \propto T^{(\nu-2)/2}\delta \phi$ goes to zero. Therefore, we conclude that the kinetic and potential energies diverge  near the big brake event but the perturbed exotic field decays to zero and  $\theta$ as well. Our overall conclusion is that  in order to  really understand  future-like singularities it  is fundamental to  explore classical perturbations   near background solutions, and when doing it is physically useful approach such cosmological scenarios  from the viewpoint of a microscopic model  such as the exotic quintessence model presented above.

We end this section by looking at the behavior of  the intrinsic entropy perturbation associated with the exotic quintessence field \cite{MaBer}.  Since  the intrinsic entropy perturbation of  matter is a gauge-invariant variable \cite{Bar}, one can use this quantity as a useful approach to investigate the behavior of the exotic quintessence model near the singular event. Such a quantity can be written in terms of nonadiabatic pressure perturbations by using Eq. (\ref{Ga}) as $\Gamma=\delta p_{\rm nad}/p$.  We focus on the first case associated with the choice $\al>0$.  Inserting Eqs. (\ref{det1}), (\ref{pet1}), and (\ref{c2a}) along with  Eqs. (\ref{solud2}) and (\ref{solup2}) into Eq.(\ref{Ga}),  we arrive at  the leading terms in the intrinsic entropy perturbation,

\bn{Ga2}
 (\nu-1)\Gamma=B (2-\nu)+\sqrt{3}c_{0}A (2-\nu) T^{\frac{(5-4\nu )}{3}}.
\ee
Equation (\ref{Ga2}) tells us that it goes to a constant as $T \rightarrow 0$  only for $\nu \in (1, \nu_{c})$ with $\nu_{c}=5/4$, whereas it remains unbounded for other values of $\nu$.  Inserting Eqs. (\ref{det1}), (\ref{pet1}), (\ref{c2a}),  (\ref{solud3}), and (\ref{pertup}) into Eq. (\ref{Ga}), the intrinsic  entropy perturbation is given by 
\bn{Ga3}
\Gamma=-2\sqrt{3}c_{0}A (2-\nu) T^{\frac{(1-2\nu )}{3}},
\ee
in the second case with $\al<0$. It is  therefore clear from this expression that  the aforesaid physical magnitude becomes divergent  near the big brake for all values of $\nu \in (1,2)$. 


\section{summary}
We have considered  four-dimensional cosmological solutions in which a flat  FRW universe exhibits a singular event at a finite time called a big brake, characterized by  a nonzero finite scalar factor, vanishing Hubble rate, and infinite acceleration.  We have shown that this cosmic scenario seems to appear when the universe is filled with two interacting components: one serves as matter, whereas the other represents a variable vacuum energy; the exchange of energy between these components  corresponds to a nonlinear function of the total energy density and its first derivative. A byproduct of this proposal is that the effective model admits an equation of state corresponding to a modified Chaplyigin gas model, depending  on the sing of a certain parameter. In solving the field equation, we applied the source equation approach because it is a useful method to reconstruct the partial energy densities, pressure, total density, and barotropic index in terms of the scale factor.  Later, we demonstrated how two interacting dark components can be mapped into an exotic quintessence (microscopic) model, which is characterized by the lack of a Lagrangian function but it can be presented as a model coming from an energy-momentum tensor.   The latter tensor is  reminiscent of a generalized scalar field theory provided the potential term can also depend on the kinetic energy.
In addition, we obtained the covariant  equation of motion for the quintessence field and showed that additional kinetic contributions appeared. 

We have examined several physical outcomes when the background solutions associated with the big brake singularity are perturbed. In doing so, we performed our analysis within the  Newtonian gauge  provided, the physical observers are attached to  the unperturbed metric so they measure both the gravitational and the velocity fields.  We have examined only the scalar mode of the metric and included entropy perturbation in the perturbed pressure. To be more precise, we have considered a general relation between pressure perturbations and density perturbations, and by doing so we  took into account the adiabatic and entropy components in  pressure perturbations. We obtained and solved a master equation for the gravitational Newtonian potential. Keeping the leading terms in the aforesaid equation, we showed that  the general solution involves the Bessel functions of  the first and second kind, which near the singular event can be written as power laws. Indeed,  we found that the Newtonian potential leads to a constant plus additional regular terms, so  it does not blow up. After that, we solved a first-order equation for the exotic quintessence field using the $0-0$ perturbed Einstein equation and employing the perturbed exotic quintessence density. We studied two different cases depending on the effective equation of state associated with the exotic quintessence field. For the $\al>0$ case, we have found that the perturbed exotic field cannot grow without limit near the singularity; in fact, it is exponentially suppressed, assuring that the  perturbation decays to zero near the big brake. We calculated other physical magnitudes; for instance, the divergence $\theta \propto T^{(2-\nu)/2}\delta \phi$ also goes to zero whereas the perturbed pressure and contrast density both become divergent near the singularity. The other case corresponding to $\al<0$ led to a perturbed exotic quintessence field that decays exponentially and the perturbed pressure is fully regular for certain values of the model parameter $\nu$. However, the contrast density  blows up. Interestingly enough,  the perturbed exotic field and the divergence of the exotic field's velocity both vanish near the big brake event. Regardless of the fact that the kinetic energy and  potential explode near the singular event, the perturbed exotic field can be controlled which means that this solution is classically stable. Such a finding showed us the critical importance of considering a microscopic model for  studying future singularities like the big brake event and the key role played by the perturbation, which helped us to explore the stability of the  solutions.  

We have shown that the intrinsic entropy perturbation  can remain bounded for certain values of $\nu$ near the big brake singularity when $\al>0$; however,  it  magnitude cannot be controlled in the $\al<0$ case, growing without limit near the singular event.  It is worthwhile to point out  that both cases are not physically equivalent at the perturbative level given the signature displayed by the intrinsic entropy perturbation.

\acknowledgments

L.P.C. thanks  U.B.A under Project No. 20020100100147 and CONICET under Project PIP 114-201101-00317.  M.G.R.  is supported by  Conselho Nacional de Desenvolvimento Cient\'ifico e Tecnol\'ogico (CNPq), Brazil.


\end{document}